# Promoting Saving for College Through Data Science

Fernando Diaz
Office of the Illinois State Treasurer
Chicago, IL, USA
FDiaz@illinoistreasurer.gov

Natnaell Mammo
Civis Analytics
Washington, DC, USA
nmammo@civisanalytics.com

## ABSTRACT

The cost of attending college has been steadily rising and in 10 years is estimated to reach $140,000 for a 4-year public university[1]. Recent surveys estimate just over half of US families are saving for college[2]. State-operated 529 college savings plans are an effective way for families to plan and save for future college costs, but only 3% of families currently use them[3].

The Office of the Illinois State Treasurer (Treasurer) administers two 529 plans to help its residents save for college. In order to increase the number of families saving for college, the Treasurer and Civis Analytics used data science techniques to identify the people most likely to sign up for a college savings plan. In this paper, we will discuss the use of person matching to join accountholder data from the Treasurer to the Civis National File, as well as the use of lookalike modeling to identify new potential signups. In order to avoid reinforcing existing demographic imbalances in who saves for college, the lookalike models used were ensured to be racially and economically balanced. We will also discuss how these new signup targets were then individually served digital ads to encourage opening college savings accounts.

## 1. MOTIVATION

529 college savings plans are tax-advantaged plans designed to encourage saving for future college costs. Withdrawals from these plans that are used for qualified higher education expenses are exempt from federal and most state taxes. 529 plans can be opened directly through the plan manager or through a financial advisor. The Office of the Illinois State Treasurer manages both an advisor and direct plan.

To make college an attainable goal for people of all income levels, the Illinois State Treasurer sought to promote its direct signup college savings plan in order to increase participation among lower and middle income households, many of which may not have a financial advisor. The first step to increasing this participation was learning more about current account holders and then use those findings to identify new potential signups.

### 1.1. CURRENT SIGNUP COUNTS

The Illinois State Treasurer administers two 529 plans that primarily serve Illinois residents but is open to all U.S. residents who are 18 years of age or older. The following counts are from the Illinois State Treasurer's accountholder data.

| Metric | Count |
|---|---|
| Active 529 Accounts | 366,078 |
| Direct 529 Accounts | 187,836 |
| New Accounts Opened per Month since 2012 | 3,250 |
| Account Holders that are Illinois Residents | 84% |

## 2. OUR APPROACH

Using the Treasurer's accountholder data, the Civis National File, publicly available survey data, and predictive modeling techniques, Civis Analytics identified the people most likely to open a college savings account. Our goals were:

- Create a detailed demographic profile of current accountholders by joining the Treasurer and Civis data
- Use this merged dataset to train a classifier to identify other Illinois residents who might sign up for an account
- Create a list of signup targets that is representative of the Illinois population
- Serve these targets digital ads and observe their signup rates

### 2.1. DATA SOURCES

Civis used accountholder data from the Treasurer that contained the accountholder's name, address, and details of the chosen investment plan. Civis also used proprietary data compiled from both commercial and public sources. The public datasets are aggregated at the census tract and block level and originate from American Community Surveys, FBI Crime Data, County Health Rankings, and several other publicly available sources. This aggregated data is combined with individual-level consumer data which contains additional demographic information and spending habits. This dataset of both neighborhood and individual-level variables provides a complete profile for each person. When combined with the State Treasurer's accountholder data, we can gain insight on the population that is currently enrolled in college savings plans.

## 2.2. METHODS

The modeling steps of this workflow primarily involved two steps, first joining the accountholder data with the consumer file, and second, using machine learning classifiers to rank order the best targets for outreach.

**Creating a Merged Dataset**

The first step was for Civis to join the State Treasurer data with the consumer file. If there was a shared unique ID between the datasets, then this would be trivial, but since such a key does not exist, we needed to be able to identify which two records referred to the same individual based off their name, address, and additional contact information. There are open-source libraries that provide this type of record linkage, but the performance of these libraries is more suited for datasets on the scale of tens or hundreds of thousands of records. Since we sought to match the holders of almost 500,000 accounts against the more than 9 million adult residents in Illinois, these libraries would not suffice.

*Step 1: Reduce the number of pairs to compare using blocking*

Blocking involves creating a set of candidate matches for each input record based on very loose matching criteria designed to retain any candidate matches that are remotely plausible given their textual and phonetic similarity to the input. This initial filtering dramatically reduces the search space for the second stage of the algorithm so that it can run efficiently, without sacrificing accuracy. Civis performs blocking by defining a set of nine "token types" that determine whether a given record should be included in the candidate set for an input record. For example, one token type specifies that records with the same city and state, and a phonetically similar name, should always be included in the candidate set. Civis uses an Amazon Dynamo DB fast-access key-value store to efficiently implement this initial blocking step. For each input record, the search space can be reduced from hundreds of millions of records down to an average of less than 40 in under 10 milliseconds.

*Step 2: Identifying the most likely pair of matches*

In the inference step, each pair of candidate matches is evaluated to determine the likelihood that the two records correspond to the same person. This likelihood is estimated based on observed features of the two records, such as whether names, addresses, phone numbers, birth dates, and email addresses match completely or in part, how common the individuals' names are, and approximate geographical matching features. The Civis file record with the highest match likelihood for a given input record is returned as the best match, and the user can then use the match likelihood scores to rank records according to the quality of the match found for each.

With our approach to matching the two datasets, we were able to match 75% of the accountholders to our consumer data. This merged dataset provided the Treasurer with insight on the audiences that were saving for college. This detailed profile extended beyond just a few basic demographic descriptors and included information about the person's household and their neighborhood. This in-depth description allowed Civis to identify people who looked similar to the current accountholders and would benefit from a 529 plan but had yet to sign up for an account themselves.

- 65% of account holders have children in the household.
- 70% of account holders live in suburban zip codes, many of whom are in the "collar counties" surrounding Chicago.
- 45% live in highly educated census tracts.
- 43% are between 45 and 59 years-old.

**Lookalike Modeling**

To identify residents who appeared similar to current accountholders and would therefore most likely sign up for an account, we constructed a training set where the dependent variable was whether the individual has a college savings account. The training set was comprised of all individuals from the accountholder dataset as the positive cases, as well as an equal number of random individuals from the Civis National File who did not match and presumably do not have an account. This resulted in a 50/50 split of positive and negative cases for the training set.

While lookalike modeling can increase enrollment by very efficient targeting and outreach, it can also perpetuate demographic imbalances that are currently reflected in the accountholder dataset. Specifically, middle income households and Hispanic and African-American households are not proportionally represented in the accountholder data when compared to the general Illinois population. Since the classifier is trained to identify targets that are similar to current accountholders, these underserved communities would very likely not appear in a list of top targets. To avoid reinforcing this already existing skew in the data, we trained five different classifiers, one for African-Americans, Asians, Hispanics, Native Americans, and Whites. Then the top targets from each classifier were proportionally pulled to construct a list that was representative of the general population.

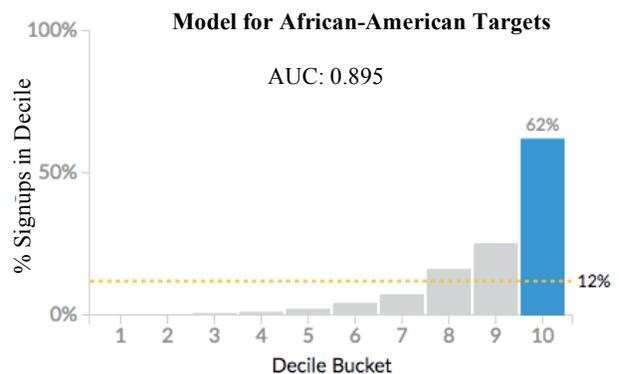

**Model for African-American Targets**
AUC: 0.895

When deciding the type of predictive classifier to train with, we had to account for the wide dimensions of our training set, which contained over 600 descriptive variables. We decided that sparse logistic classifiers best suited our needs because of its ability to avoid overfitting, be computationally feasible, and maintain interpretability. One important way we evaluated the model's rank ordering ability was through decile plots. The decile plots used out-of-sample scores from k-fold cross validation (k=4). The example above demonstrates that 62% of the top decile of African-American individuals that the model selected as being likely to open an account, in fact did sign up. This is more than five times as effective than a random sample which would return 12% of signups. The appendix shows decile plots for each of the other models. Four of the five models showed strong rank ordering; the model for Native American targets did not rank as strongly due to the smaller sample size in the training set. Since Lasso regression performs variable selection, we inspected the variables that were selected and found that the most predictive features were the probability of being a parent, probability of being a homeowner, and the average income of the household and census tract.

**Using Predictive Modeling for Digital Outreach**

After constructing a list of the top 720,000 targets, the State Treasurer and Civis partnered with a digital ad company in order to serve these individuals digital ads encouraging saving for college.

## 2.3. RESULTS

During and after the digital ad campaign, we were able to measure the effectiveness of our lookalike modeling by observing how many targets opened an account.

- Targets were 4.4 times more likely to open an account than non-targets.
- The number of total account signups for the first quarter of 2017 is 23% higher than the previous year.
- The number of account signups for the direct 529 plan is 26% higher than the previous year.
- We found that 18% of the new accounts in 2017 were opened by a target.
- We decided to replicate this count at the household level since one member of a family may be a target but a different family member may actually open the account, as would likely be the case for married couples. 27% of new accounts were opened by a person in the same household as a target.

## 3. CONCLUSION

Our research with the Treasurer's Office showed that traditional or industry-standard messaging about saving for college only has a marginal impact on sign-ups. Market-segmented message testing is vital to ensuring effectiveness, particularly as organizations look to appeal to non-traditional participants. State governments that have limited budgets must use innovative and cost effective ways to spread information about college savings plans, particularly amongst "non-traditional" college savers. The person level modeling and targeting employed by the Illinois State Treasurer and Civis Analytics effectively increased participation by maximizing the efficiency of its outreach and each advertising dollar. This approach of joining multiple sources of data to enrich usually isolated datasets and using lookalike modeling can be replicated by any organization seeking to better understand its target audience and also encourage new participants in its public services.

## 4. APPENDIX

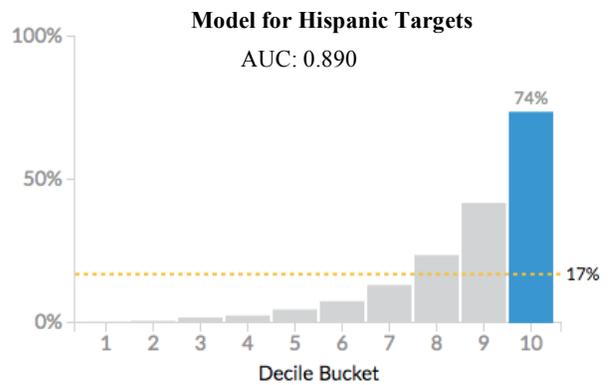

Model for Hispanic Targets
AUC: 0.890

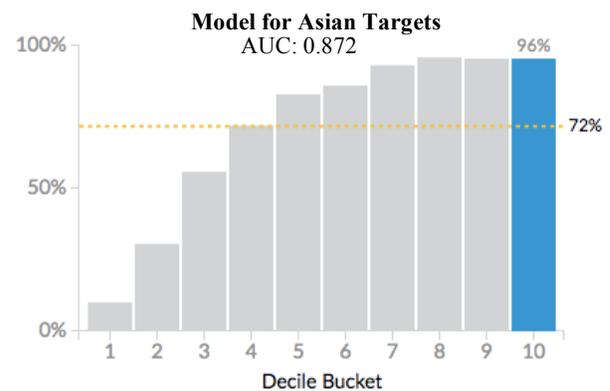

Model for Asian Targets
AUC: 0.872

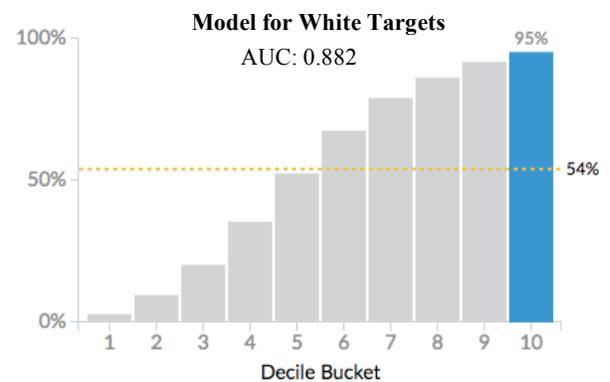

Model for White Targets
AUC: 0.882

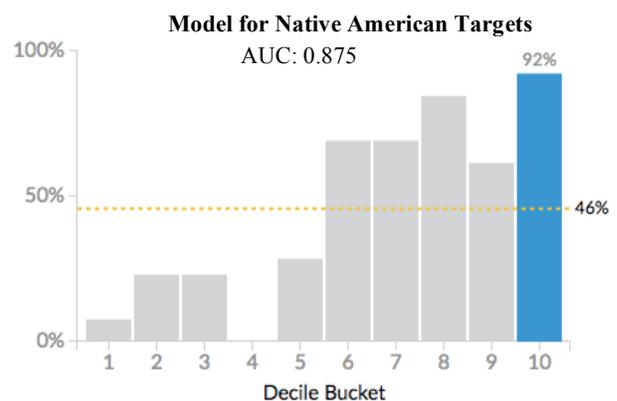

Model for Native American Targets
AUC: 0.875

## 5.REFERENCES